# A Federated Architecture for Sector-Led AI Governance: Lessons from India

Avinash Agarwal[1][1] and Manisha J. Nene[2]

[1]Department of Telecommunications, Ministry of Communications, Sanchar Bhawan, New Delhi, India

[2]Defence Institute of Advanced Technology, Ministry of Defence, Girinagar, Pune, India

March 2, 2026

**Abstract**

**Purpose** - India has adopted a vertical, sector-led AI governance strategy. While promoting innovation, such a light-touch approach risks policy fragmentation. This paper proposes a cohesive 'whole-of-government' architecture to mitigate these risks and connect policy goals with a practical implementation plan.

**Design/methodology/approach** - The paper applies an established five-layer conceptual framework to the Indian context. First, it constructs a national architecture for overall governance. Second, it uses a detailed case study on AI incident management to validate and demonstrate the architecture's practical utility in designing a specific, operational system.

**Findings** - The paper develops two actionable architectures. The primary model assigns clear governance roles to India's key institutions. The second is a detailed, federated architecture for national AI Incident Management. It addresses the data silo problem by using a common national standard that allows sector-specific data collection while facilitating crosssectoral analysis.

**Originality/value** - This paper proposes a detailed operational architecture for India's 'whole-of-government' approach to AI. It offers a globally relevant template for any nation pursuing a sector-led governance model, providing a clear implementation plan. Furthermore, the proposed federated architecture demonstrates how adopting common standards can enable cross-border data aggregation and global sectoral risk analysis without centralising control.

**Practical implications** - The proposed architectures offer a clear and predictable roadmap for India's policymakers, regulators, and industry to accelerate the national AI governance agenda.

**Social implications** - By providing a systematic path from policy to practice, the architecture builds public trust. This structured approach ensures accountability and aligns AI development with societal values.

## 1 Introduction

Artificial Intelligence (AI) has emerged as a transformative general-purpose technology, creating significant opportunities and complex governance challenges for nations worldwide. A key challenge, widely recognised in the global AI Governance (AIG) literature, is the 'policy-to-practice gap': the difficulty of translating high-level ethical principles into practical, verifiable actions on the ground [Birkstedt et al., 2023, Sadek et al., 2025]. Without a clear and cohesive implementation architecture, even well-defined national strategies risk policy fragmentation and inconsistent standards. This can lead to tangible negative consequences, such as the deployment of biased AI systems in critical sectors or unclear liability when AI failures occur.

This implementation challenge is particularly acute for countries like India that have deliberately chosen a flexible, sector-led governance model over a single, overarching AI Act [IndiaAI, 2025]. While this approach is pro-innovation, it inherently faces the challenge of 'structured decentralisation', the need to ensure coherence across a diverse set of actors. This operational gap is particularly evident in specific domains. For example, while the need for AI incident reporting is widely accepted, the practical mechanisms for a federated, multi-repository system remain undefined, risking underreporting in critical sectors such as power and telecommunications [Agarwal and Nene, 2024a]. Similarly, while national standards for AI fairness have been developed [Telecommunication Engineering Centre, 2023], the pathway to a trusted, voluntary compliance ecosystem that can build market credibility for startups is still in the process of evolving. These examples, common to many jurisdictions pursuing a vertical model, underscore the urgent need for a cohesive implementation architecture.

---

[1] Corresponding author: avinash.70@gov.in

This paper aims to bridge this specific implementation gap. It applies a previously published and validated five-layer framework for AI governance [Agarwal and Nene, 2025] to the specific Indian context, translating the government's strategic vision into an actionable, multi-stakeholder architecture. The framework provides a structured pathway from high-level laws and policies (Layer 1) through standards and assessment procedures (Layers 2 and 3) to the development of tools and the creation of a trusted ecosystem (Layers 4 and 5).

The primary impact of this work is to provide a clear and predictable roadmap for India's policymakers. More broadly, it offers a globally relevant template for any country pursuing a sectorled, vertical approach to AI governance. By providing a structured, multi-layered architecture, this paper moves the national conversation from a debate over principles to a focused discussion on implementation. It offers a model that preserves an agile and pro-innovation stance while creating the robust structures necessary to build public trust, ensure accountability, and manage risk effectively.

The specific contributions of this paper are threefold:

1. It presents a detailed, operational architecture for India's *whole-of-government* approach to AI governance.

2. It provides two concrete blueprints derived from this architecture: a comprehensive plan for overall national governance and a deep-dive case study on a national AI incident management system.

3. It demonstrates how a federated, multi-layered model can effectively balance central coordination with decentralised, sector-specific implementation, providing a scalable protocol for future international cooperation and global incident data aggregation.

The remainder of this paper is structured as follows. Section 2 reviews the relevant literature on AI governance. Section 3 outlines the research methodology and briefly recaps the analytical framework used for this analysis. Section 4 presents the India-specific architecture for overall AI governance, detailing the roles and responsibilities of key stakeholders. Section 5 applies this architecture to the specific challenge of AI incident management, presenting a detailed operational model. Section 6 discusses the policy implications of this work, and finally Section 7 offers concluding remarks.

## 2 Literature Review

This section reviews the literature on AI governance to situate the paper's contribution. It outlines the global landscape of AI governance models and then examines India's policy approach, identifying the research gap this paper aims to address.

### 2.1 The Global Landscape of AI Governance

Diverse approaches characterise the global AI governance landscape, as nations worldwide seek to harness the benefits of this transformative technology while mitigating its inherent risks [Taeihagh, 2021]. Some countries are pursuing a comprehensive, regulation-first model, best exemplified by the European Union's AI Act [Parliament, 2024]. Others, such as the United States, have adopted a decentralised, sector-specific approach complemented by soft-law instruments such as the NIST AI Risk Management Framework [Tabassi, 2023]. At the multilateral level, organisations such as the Organisation for Economic Co-operation and Development (OECD) have provided foundational 'Principles on AI' [OECD, 2019], while the United Nations Educational, Scientific and Cultural Organisation (UNESCO) has offered a 'Recommendation on the Ethics of Artificial Intelligence' [UNESCO, 2022]. This has led to a convergence on high-level principles, but the path to implementation remains fragmented. Scholars argue that a single, overarching global institution is unlikely to be feasible, suggesting the more realistic path forward lies in strengthening the existing complex of international regimes [Roberts et al., 2024].

Broadly, two distinct models are emerging globally within a 'nascent regime in a fragmented landscape' [Schmitt, 2022]. The first is the comprehensive, horizontal approach, exemplified by a single, overarching AI Act. The second and more common model is the vertical, sector-specific approach. This model, adopted by nations such as the United States, the United Kingdom, and India, avoids a single AI law and instead empowers existing sectoral regulators to adapt their legal frameworks to the challenges posed by AI in their domains [Smuha, 2021, Agarwal and Nene, 2026, Kashefi et al., 2024].

This policy-to-practice gap is a well-recognised and critical challenge in the global AI Governance (AIG) literature. Systematic reviews of the field highlight a significant lack of progress in operationalising AIG processes [Birkstedt et al., 2023]. A primary reason is that many highlevel ethical guidelines remain too abstract to provide concrete direction during the design and development of AI systems [Sadek et al., 2025]. Furthermore, the mere existence of these principles is far from sufficient to ensure the trustworthiness of AI systems in practice, creating significant hurdles for their operationalisation [Zhu et al., 2021]. Consequently, many organisations report struggling to translate their Responsible AI commitments into meaningful and verifiable actions [Akbarighatar, 2025]. The literature, therefore, has thoroughly documented the *what* (the principles to be followed) but has yet to provide a clear answer for the *how*: the specific, repeatable architecture needed for verifiable implementation.

## 2.2 The AI Governance Landscape in India

India's pro-innovation approach to AI governance prioritises technological growth while maintaining a focus on responsible, risk-based oversight. The foundational vision, established in NITI Aayog's 2018 'National Strategy for Artificial Intelligence', is that of *#AIforAll*, positioning AI as a primary driver for inclusive and social growth [NITI Aayog, 2018]. This vision was further refined with a clear focus on building public trust through principles for *Responsible AI for All* [NITI Aayog, 2021]. This strategy deliberately embraces a *light-touch* regulatory model, a stance confirmed in Parliament in April 2023 when the government stated it was not considering a single, overarching law [The Economic Times, 2023]. Instead, the approach is to build on India's existing and robust legal framework, including overarching statutes such as the Information Technology Act, the Digital Personal Data Protection Act, and the Copyright Act, as well as domain-specific laws such as the Telecommunications Act, to create an agile and adaptive governance system. To orchestrate this decentralised system, the government is committed to a *whole-of-government* approach [Banzal and Agarwal, 2025], a direction further detailed in a draft report from the 'Subcommittee on AI Governance and Guidelines Development' [IndiaAI, 2025] and the official 'India AI Governance Guidelines' [MeitY, 2025].

In the Indian context, the operational gap identified in the literature manifests in specific domains, as introduced in Section 1. Current scholarship confirms that key governance functions, such as incident reporting in critical infrastructure and fairness certification for the startup ecosystem, lack the necessary institutional mechanisms for effective implementation [Agarwal and Nene, 2024a, Agarwal et al., 2023]. These examples underscore the urgent need for a cohesive implementation plan in India, which this paper aims to provide.

# 3 Methodology

This study employs an applied conceptual framework methodology. This approach is specifically selected to address the 'policy-to-practice' gap by using an established theoretical model as an analytical lens to structure a complex governance problem and derive a context-specific architectural solution. The research design follows three distinct stages.

## 3.1 Data Sources and Thematic Analysis

The data sources for this paper are a curated set of high-level policy and strategic documents, as well as foundational academic literature. These include the European Union's AI Act, the NIST AI Risk

Management Framework, and key Indian policy documents, including NITI Aayog's National Strategy for Artificial Intelligence and MeitY's India AI Governance Guidelines. Thematic analysis of these sources revealed the core problem: a gap between high-level principles and practical implementation. This gap is especially stark in vertical, sector-led governance models such as India's, where 'structured decentralisation' remains a key challenge.

## 3.2 The Analytical Framework: A 5-Layer Model for AI Governance

The second stage involved selecting the analytical framework. While various governance models exist, the five-layer framework for AI governance [Agarwal and Nene, 2025] was chosen for three critical reasons. First, unlike models that focus solely on high-level principles [Birkstedt et al., 2023], it provides a comprehensive, whole-of-stakeholder architecture designed to bridge this 'policy-practice gap'. Its key strength is that it outlines a systematic pathway from high-level principles to verifiable, on-the-ground implementation. The concept of layered governance, where different architectural levels address distinct problems, is a well-established approach in technology policy [Gasser and Almeida, 2017]. Second, the framework has been validated through a peer-reviewed publication in this journal (Transforming Government: People, Process and Policy), establishing its academic credibility. Third, and most significantly, the framework has been recognised in India's official policy discourse, having been cited in the 'India AI Governance Guidelines' released by the Ministry of Electronics & Information Technology (MeitY) as a reference for mapping key agencies and functions [MeitY, 2025]. This makes it the most relevant and empirically grounded tool for analyzing the Indian context.

This established framework organises the governance ecosystem into a logical cascade of five distinct layers:

1. **Laws, Regulations & Policies**: This foundational layer establishes the overarching legal and ethical mandates. It is the domain of government and public institutions that defines the national strategy for AI.

2. **Standards**: This layer translates the broad principles from Layer 1 into specific, technical, and measurable requirements. This is the primary function of national and international standards-setting bodies.

3. **Standardised Assessment Procedures**: This layer defines the consistent and repeatable methodologies for testing and verifying compliance against the standards established in Layer 2.

4. **Standardised Tools & Metrics**: This layer comprises the practical software tools, libraries, and quantifiable metrics used to execute the assessment procedures from Layer 3, enabling scalable and objective evaluation.

5. **Voluntary Compliance Ecosystem**: The final layer facilitates building public trust. It includes mechanisms for demonstrating compliance, such as third-party certification and transparency reporting, and provides crucial feedback loops to the upper layers.

The framework operates as a dynamic system built upon a logical cascade. Figure 1 illustrates this primary cascade, showing how it systematically translates a policy objective into a trusted AI system through the five layers. While the figure shows the core top-down flow, the framework also incorporates a crucial feedback loop. Data and real-world evidence from the operational layers flow upward to inform and refine policies and standards in the higher layers. This ensures the system remains adaptive. A potential limitation of this framework is its reliance on extensive inter-agency coordination for effective operation. The architecture proposed in this paper addresses this challenge by clearly delineating roles and responsibilities.

## 3.3 Architectural Construction and Case Study Validation

The final stage involved architectural construction and validation. Using the five-layer framework as a structural guide, this study mapped key institutional actors and governance functions within the Indian ecosystem, as identified through thematic analysis, to the corresponding layers. This deductive process culminated in the development of a comprehensive AI governance architecture, presented in the next section.

To validate the proposed architecture and demonstrate its practical utility, an illustrative case study on AI incident management was developed. A single-case methodology is appropriate here because the research goal is deductive: not to inductively build a new theory from multiple data points, but to demonstrate the architecture's analytical power in structuring a complex, real-world problem and designing a specific, operational system. The incident management case was chosen because it is a timely and critical policy challenge that has been explicitly identified as a priority in the recent national guidelines [MeitY, 2025], making it a perfect exemplar for the kind of multistakeholder coordination the architecture provides.

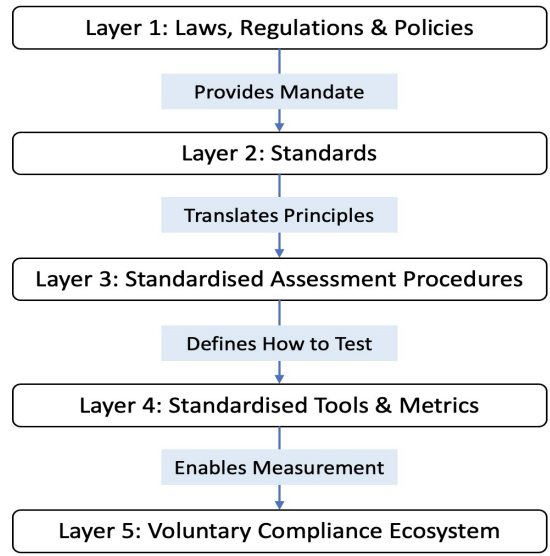

*Figure 1: A Cascade Model for AI Governance Implementation (Source: Adapted from [Agarwal and Nene, 2025]). Note: This simplified model illustrates the logical flow from high-level policy to ecosystem-level trust.*

# 4 An India-Specific Architecture for AI Governance

As established in Section 2, India's vertical, sector-led approach requires a unifying architecture to ensure coherence and prevent fragmentation. This section proposes such an architecture. It translates the abstract logic of the five-layer framework into a concrete operational model for India, mapping key governance functions to the specific Responsibility Centres accountable for their execution. This provides a clear, whole-of-government roadmap for translating a national principle into a trusted AI system in the market, as detailed in Table 1.

The proposed architecture assigns clear responsibilities at each governance layer. While built upon the framework's top-down cascade from policy to implementation, the architecture is designed as a dynamic system. As detailed below, guidance and standards flow downwards, while data and evidence flow upwards, ensuring the system remains adaptive.

At the apex (Layer 1), the high-level AI Governance Group, supported by its Technical Secretariat, provides the central vision, sets the national risk-based approach, and ensures interministerial coordination [MeitY, 2025]. This body guides the Line Ministries and Sectoral Regulators in applying and adapting existing legal instruments to mandate risk-graded compliance for critical applications. This structure formally addresses the risk of fragmentation by creating a central point for guidance while preserving the domain-specific expertise of individual regulators.

To ensure consistency across federated sectors, Layer 2 emphasises standardisation as the common language of the entire ecosystem. National standards bodies, such as the Bureau of Indian Standards (BIS) and the Telecommunication Engineering Centre (TEC), are responsible for creating uniform definitions and technical benchmarks. The voluntary 'Standard for Fairness Assessment and Rating of Artificial Intelligence Systems' developed by TEC is a key example of this sector-led approach [Telecommunication Engineering Centre, 2023]. Importantly, this layer also underscores the need to

">
5
">
*This is the author's accepted manuscript of the article published as: Avinash Agarwal, Manisha J. Nene, "A federated architecture for sector-led AI governance: lessons from India", Transforming Government: People, Process and Policy, 2026,*
*https://doi.org/10.1108/TG-09-2025-0310*

align domestic standards with global best practices from organisations such as the International Telecommunication Union (ITU) and the International Organization for Standardization (ISO). Such alignment ensures interoperability, facilitates cross-border trade, and strengthens the global competitiveness of India's AI industry.

*Table 1: An India-Specific Architecture for AI Governance (Source: Authors' own work, integrating institutional entities from [MeitY, 2025])*

| # | Layer | Responsibility Centres | Core Responsibilities |
|---|---|---|---|
| 1 | Laws, Regulations & Policies | AI Governance Group | High-Level Policy: Set national riskbased approach |
| | | | Overall supervision, guidance, and interministerial coordination |
| | | Line Ministries, Sectoral Regulators | Domain-Specific Rules: Binding rules under existing Acts |
| | | | Mandate risk-graded compliance for critical applications |
| | | | Domain-specific grievance handling |
| | | Technical Secretariat (MeitY, AISI) | Support to the Apex AI Group |
| 2 | Standards | Standards Bodies (BIS, TEC) | Develop National Standards |
| | | | Definitions, taxonomies, requirements |
| | | | Align with global bodies (ITU, ISO/IEC, IEEE) |
| 3 | Standardised Assessment Procedures | Academia, Industry | Standardise assessment (test) procedures (e.g., Auditing guidelines, Incident verification protocols) |
| | | Standards Bodies (BIS, TEC, AISI) | |
| | | Technical Secretariat | Topics identification and progress monitoring |
| 4 | Standardised Tools & Metrics | Academia, Startups | Develop standards-aligned assessment tools, APIs, and libraries |
| | | MeitY, DST, etc. (Funders) | Funding to encourage innovation. |
| | | Technical Secretariat | Topics identification and progress monitoring |
| 5 | Voluntary Compliance Ecosystem | Industry, Developers, Deployers | Self declarations, transparency disclosures |
| | | Neutral Third-Parties (Academic Labs) | Network of Assessment Bodies (CABs) and Certification Bodies |
| | | MeitY, DST, etc. (Funders) | Funding to encourage entrepreneurship |
| | | Technical Secretariat | Drive awareness campaigns |

At Layer 3, a collaborative ecosystem of expert bodies can develop consistent test procedures, such as auditing procedures and incident verification protocols. It includes key institutions like the AISI, an institutional model now being adopted by several nations to address complex safety risks [Hwang and Kwon, 2025], alongside academia, industry, and standards bodies. It provides a transparent and repeatable methodology for verification, with the Technical Secretariat playing a key role in identifying topics and monitoring progress.

This, in turn, supports the development of practical tools and metrics in Layer 4. At this stage, India's innovation ecosystem, particularly academia and startups, can create standards-aligned tools and APIs that make large-scale compliance feasible. Key bodies, such as the Ministry of Electronics and Information Technology (MeitY) and the Department of Science and Technology (DST), may support this work with government funding. The government acts as a catalyst, funding the development of these tools to build capacity in AI assurance.

Finally, Layer 5 relies on a suite of accountability tools to build trust in a light-touch regulatory environment. Mechanisms such as self-declarations [Arnold et al., 2019], transparency disclosures [OECD, 2019], or voluntary third-party certifications [Blösser and Weihrauch, 2024], serve as key mechanisms for building a trustworthy AI ecosystem. This flexible approach enables innovators, particularly medium and small enterprises, to build market credibility and demonstrate due diligence in ways suited to their scale. This final layer also enables the crucial upward feedback loop, where real-world data and learnings are channelled back to the apex body, ensuring the entire system remains adaptive, evidence-based, and grounded in public trust.

## 5 An AI Incident Management Case Study

While the governance architecture described in the previous section is broadly applicable across various AI risks, this section demonstrates its practical utility by applying it to the specific and complex challenge of AI incident management. This domain serves as an ideal illustrative test case because it is critical for building public trust and requires precisely the kind of multi-stakeholder coordination that a whole-of-government approach demands. Furthermore, the need to establish a national AI incident database is a specific recommendation in recent government policy drafts, making this a timely and relevant challenge [IndiaAI, 2025, MeitY, 2025].

A central question in designing such a system is whether to create a single, centralised database or a federated system of multiple, sector-specific ones. While a single database offers the theoretical benefit of cross-sectoral learning, evidence from existing global incident databases reveals a significant flaw: general-purpose repositories are consistently overwhelmed by incidents from high-visibility domains such as autonomous vehicles and social media. Consequently, critical but less public-facing sectors such as telecom and energy remain severely underreported [Agarwal and Nene, 2024a, Avinash and Manisha, 2024].

This paper, therefore, proposes a federated model of sector-specific AI incident databases as a more effective solution for India. This approach aligns with India's vertical governance structure, empowering sectoral regulators to mandate and incentivise participation. Crucially, this federated model avoids creating data silos, a common challenge in decentralised governance, through a key architectural choice: a single, nationally mandated standard for the database schema. When every repository collects data in the same format, a central body such as the AISI can aggregate the anonymised information. This approach offers the focused data collection of a decentralised system, along with the holistic insights characteristic of a centralised system.

Figure 2 visualises the architecture for operationalising this federated model. It represents a dynamic, interconnected system rather than a static hierarchy, showing how data, guidance, and learning flow across different responsibility centres.

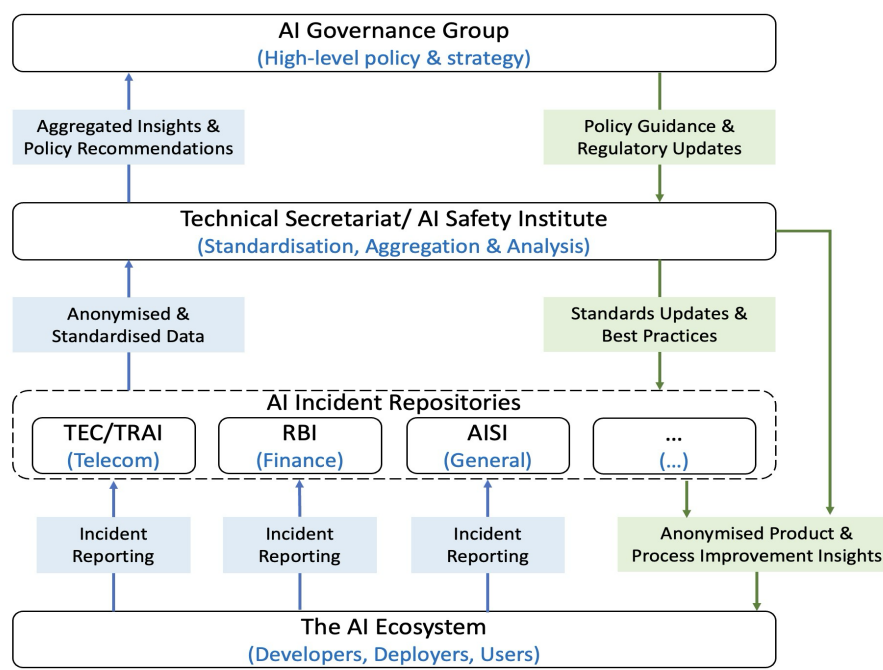

*Figure 2: An Architecture for a Federated AI Incident Management System in India (Source: Authors' own work)*

As illustrated in Figure 2, this system is distinct from the general governance architecture in its operational focus. While the same institutions are involved, their roles are tailored to the specific lifecycle of incident management.

The process begins at Layer 1 with the establishment of clear legal mandates for reporting. The architecture empowers sectoral regulators, such as the Department of Telecommunications, to use existing laws (e.g., the Telecommunications Act, 2023) to establish risk-graded reporting requirements. This allows for jurisdiction-specific implementation that is both authoritative and context-aware. For example, rules could mandate licensed service providers to report high-risk incidents occurring in core network components, while simultaneously facilitating voluntary reporting for low-risk cases. This approach, proposed for the telecommunications sector [Agarwal and Nene, 2026], can be replicated across other critical sectors, ensuring that the responsibility for risk-grading and defining critical use-cases lies with domain experts.

The reported data then flows upwards through a system governed by standardisation and common procedures at Layers 2 and 3. The interoperability of the entire federated network depends on a nationally mandated standardised schema and taxonomy (Layer 2), a step crucial for creating operational harmony and consistency across the different sectoral databases. Earlier research proposed a comprehensive schema and taxonomy after analysing the limitations of global databases [Agarwal and Nene, 2024b]. This work now serves as the foundation for the national standard on this topic, currently being finalised by TEC [Telecommunication Engineering Centre, 2025]. These standards are then operationalised through consistent procedures (Layer 3) for incident verification, root cause analysis, and secure data anonymisation, which are essential for ensuring the quality and integrity of the collected data.

The core operational activities of the federated model take place at Layer 4, which focuses on repositories, tools, and metrics. This is where the decentralised structure becomes tangible. Specific sectoral nodal bodies are designated to host and maintain their domain-specific incident repositories. For instance, TEC or TRAI could manage the telecommunications sector database, the Reserve Bank of India (RBI) could oversee the financial sector, and the AISI could maintain a general-purpose or cross-sectoral repository of AI incidents. This layer also involves developing analytical tools within the academic and startup ecosystem to monitor emerging trends, with the Technical Secretariat playing a key role in coordinating across these nodal bodies. A central body, such as the AISI, then aggregates the anonymised data from these sectoral nodes, performing the cross-sectoral analysis that provides a holistic, national-level view of AI risks.

Finally, Layer 5 ensures that the system functions as a dynamic feedback and learning ecosystem, rather than a static data repository, a key principle for building robust and trustworthy AI systems [McGregor, 2021]. This is a key differentiator from a simple reporting system. The insights generated by the Technical Secretariat and AISI are fed in two directions: upwards to the AI Governance Group to inform policy and regulatory updates, and downwards, back to the broader AI ecosystem. The anonymised trends and best practices disseminated downwards enable developers and deployers to learn from systemic failures, support data-driven improvements, advance mitigation techniques, and foster a culture of continuous learning. The funding mechanisms in this layer are targeted to encourage the development of safer AI systems. The Technical Secretariat could lead national awareness campaigns to foster a culture of responsible disclosure. This closes the loop, transforming the incident database from a static repository into a dynamic engine for a safer AI environment for India.

# 6   Discussion and Policy Implications

The primary contribution of this work is an operational architecture designed to address the implementation challenges inherent in a vertical, sector-led AI governance model. As India has deliberately chosen to forego a single, overarching AI Act in favor of a more agile approach that leverages existing legal frameworks, the central challenge becomes one of ensuring coherence and preventing regulatory fragmentation. The architectures presented in this paper offer a structured solution to this specific problem, demonstrating a systematic approach for whole-of-government operationalisation.

The proposed national governance architecture provides a mechanism for *structured decentralisation*. A key implication of this model is that it creates a precise and predictable relationship between the apex coordinating bodies and the various implementing agencies. The core contribution of this paper is not the definition of the layers themselves, but rather the application of this layered model to India's institutional context, identifying the key responsibility centres and mapping their specific roles and functions to each layer. This architecture translates the strategic vision of the central bodies into a coordinated set of actions. It directly addresses the risk of fragmentation by providing a clear implementation structure, thereby enhancing the predictability and stability of India's regulatory environment. A stable and predictable environment, supported by robust regulatory frameworks, is a key prerequisite for promoting AI innovation and encouraging its acceptance [Rane et al., 2024].

The case study on AI incident management serves to validate and illustrate the practical utility of this layered approach. The proposed federated model of sector-specific incident repositories offers a significant policy implication, as it directly addresses a known limitation of monolithic, generalpurpose incident databases. Academic reviews of the current landscape confirm that such databases are often overwhelmed by high-volume noise from a few domains, and a key recommendation is to move towards a more diversified and context-specific approach to incident documentation [Turri and Dzombak, 2023]. Our architectural design, which mandates a common national standard for the incident schema (Layer 2), enables this diversification. It allows domain experts at sectoral nodal bodies to manage high-quality, relevant data (Layer 4). A central body can also aggregate anonymised data from various such sectoral databases for a holistic, national-level risk assessment.

Furthermore, the operational model proposed in this paper has broader relevance beyond the Indian context. Many nations are grappling with the same choice between a horizontal AI Act and a vertical, sector-led model. The architecture detailed here provides a potential template for implementing such a vertical approach, particularly for countries facing similar challenges of balancing innovation with limited regulatory interventions and the need for more contextualised governance frameworks [Kashefi et al., 2024]. It offers a case study in balancing the flexibility of decentralised regulation with the need for national-level consistency and strategic oversight. This is especially relevant because the global landscape features diverse regulatory strategies, each reflecting the unique societal and economic context of its country, making a 'one-size-fitsall' approach ineffective [Kaliisa et al., 2025].

While the model offers a versatile template, its transferability to other countries is contingent upon certain boundary conditions. The success of this federated architecture depends on the existence of mature sectoral regulators and standards-setting bodies with sufficient technical and administrative capacity. In jurisdictions with lower institutional readiness, the decentralised nature of this model could inadvertently lead to further coordination challenges. Consequently, while the architecture provides a scalable roadmap, its implementation must be calibrated to the specific institutional maturity and state capability of the host nation.

Finally, the architectural principles proposed here hold significant potential for international AI governance. If nations adopt harmonised standards for sector-specific incident schemas, as proposed in Layer 2, it lays the groundwork for a global federated network. For instance, national telecom-specific incident repositories could feed anonymised data to international bodies such as the International Telecommunication Union (ITU). Similarly, a nodal global financial organisation could aggregate AI incident data in the financial sector from country-specific databases. This would transform isolated national insights into a global intelligence network, allowing for the rapid identification of systemic AI failures in critical sectors across borders without compromising national data sovereignty.

Looking ahead, the successful implementation of this architecture implies a focus on capacity building at each layer. A key prerequisite is strengthening the institutional capabilities of the designated responsibility centres, a challenge common to regulatory ecosystems globally [Aitken et al., 2022]. It includes empowering national standards bodies to develop timely and globally harmonised standards (Layer 2) and equipping the AISI to define robust assessment procedures (Layer 3). A practical path forward would be to prioritise the AI incident management framework as an initial implementation project. It could be operationalised within a regulatory sandbox environment, a tool that allows policymakers to better understand disruptive technologies like AI before creating binding rules [Yordanova and Bertels, 2024]. Such an approach would serve as an effective pathway for building institutional capacity while demonstrating the value of a structured, architectural approach to AI governance.

# 7 Conclusion

As India charts its own course in AI governance, opting for a flexible, *light-touch*, sector-led model over a single, overarching law, it faces a significant challenge in ensuring coherence in a decentralised system. High-level principles and well-defined institutions provide a strong foundation, but they do not, by themselves, constitute a functioning system. This paper addresses the critical gap by providing the operational architecture that connects these disparate parts into a cohesive, *wholeof-government* approach.

The primary contribution of this work is the provision of this architecture. By applying an established five-layer framework to the Indian context, this paper presents a clear and practical roadmap that moves beyond principles to the specifics of implementation. Through the comprehensive national governance architecture and the case study on AI incident management, this work demonstrates how a *whole-of-government* approach can be practically achieved, balancing the need for central strategic coordination with the necessity of decentralised, domain-specific execution. It provides a structured pathway from a policy goal to a trusted AI system in the market.

The implications of this architectural approach are significant. For domestic policymakers, this offers a predictable and stable model. It brings clarity to both regulators and industry, enabling an environment where responsible innovation can thrive. For the international community, it provides a globally relevant template for managing the inherent risks of regulatory fragmentation while serving as a foundational architecture for global interoperability. Ultimately, by providing a clear and actionable path from policy to practice, this work offers a way for India to realise its ambition: to become a global leader not just in the development of AI, but in the responsible and effective governance of it.

# References

Avinash Agarwal and Manisha J. Nene. Addressing AI Risks in Critical Infrastructure: Formalising the AI Incident Reporting Process. In *2024 IEEE International Conference on Electronics, Computing and Communication Technologies (CONECCT)*, pages 1–6, 2024a. doi: 10.1109/CONECCT62155.2024.10677312.

Avinash Agarwal and Manisha J. Nene. Standardised Schema and Taxonomy for AI Incident Databases in Critical Digital Infrastructure. In *2024 IEEE Pune Section International Conference (PuneCon)*, pages 1–6, 2024b. doi: 10.1109/PuneCon63413.2024.10895867.

Avinash Agarwal and Manisha J. Nene. A five-layer framework for AI governance: integrating regulation, standards, and certification. *Transforming Government: People, Process and Policy*, 19(3):535–555, 05 2025. ISSN 1750-6166. doi: 10.1108/TG-03-2025-0065. URL https://doi.org/10.1108/TG-03-2025-0065.

Avinash Agarwal and Manisha J. Nene. Incorporating AI incident reporting into telecommunications law and policy: Insights from India. *Computer Law & Security Review*, 60:106263, 2026. ISSN 2212-473X. doi: https://doi.org/10.1016/j.clsr.2026.106263.

Avinash Agarwal, Harsh Agarwal, and Nihaarika Agarwal. Fairness Score and process standardization: framework for fairness certification in artificial intelligence systems. *AI and Ethics*, 3 (1):267–279, 2023.

Mhairi Aitken, David Leslie, Florian Ostmann, Jacob Pratt, Helen Margetts, and Cosmina Dorobantu. Common regulatory capacity for AI. *The Alan Turing Institute*, 2022.

Pouria Akbarighatar. Operationalizing responsible AI principles through responsible AI capabilities. *AI and Ethics*, 5(2):1787–1801, 2025.

Matthew Arnold, Rachel KE Bellamy, Michael Hind, Stephanie Houde, Sameep Mehta, Aleksandra Mojsilovi´c, Ravi Nair, K Natesan Ramamurthy, Alexandra Olteanu, David Piorkowski, et al. FactSheets: Increasing trust in AI services through supplier's declarations of conformity. *IBM Journal of Research and Development*, 63(4/5):6–1, 2019.

Agarwal Avinash and Nene Manisha. Advancing Trustworthy AI for Sustainable Development: Recommendations for Standardising AI Incident Reporting. In *2024 ITU Kaleidoscope: Innovation* *and Digital Transformation for a Sustainable World (ITU K)*, pages 1–8, 2024. doi: 10.23919/ITUK62727.2024.10772925.

Preeti Banzal and Tejal Agarwal. OPSA's Role in Empowering India's AI Revolution. *Vigyan Dhara*, September 2025. ISSN 3107-6610. Accessed: 2025-09-22.

Teemu Birkstedt, Matti Minkkinen, Anushree Tandon, and Matti Mäntymäki. AI governance: themes, knowledge gaps and future agendas. *Internet Research*, 33(7):133–167, 2023.

Myrthe Blösser and Andrea Weihrauch. A consumer perspective of AI certification–the current certification landscape, consumer approval and directions for future research. *European Journal of Marketing*, 58(2):441–470, 2024.

Urs Gasser and Virgilio AF Almeida. A layered model for AI governance. *IEEE Internet Computing*, 21(6):58–62, 2017.

Seong-Ah Hwang and Hun-Yeong Kwon. A Comparative Study on the Role of AI Safety Institutes in Shaping Global AI Governance. In *2025 IEEE/ACIS 23rd International Conference on Software Engineering Research, Management and Applications (SERA)*, pages 537–542. IEEE, 2025.

IndiaAI. Report on AI Governance Guidelines Development. https://indiaai.gov.in/article/report-on-ai-governance-guidelines-development, January 2025. Accessed: 2025-09-22.

Rogers Kaliisa, Ryan Shaun Baker, Barbara Wasson, and Paul Prinsloo. The Coming but Uneven Storm: How AI Regulation Will Impact AI and Learning Analytics Research in Different Countries. *Journal of Learning Analytics*, 12(2):140–157, 2025.

Pouya Kashefi, Yasaman Kashefi, and AmirHossein Ghafouri Mirsaraei. Shaping the future of AI: balancing innovation and ethics in global regulation. *Uniform Law Review*, 29(3):524–548, 2024.

Sean McGregor. Preventing repeated real world AI failures by cataloging incidents: The AI incident database. In *Proceedings of the AAAI Conference on Artificial Intelligence*, volume 35, no. 17, pages 15458–15463, 2021.

MeitY. India AI Governance Guidelines: Enabling Safe and Trusted AI Innovation. Policy guidelines, Ministry of Electronics and Information Technology, Government of India, November 2025.

NITI Aayog. National Strategy for Artificial Intelligence. https://www.niti.gov.in/sites/default/files/2023-03/National-Strategy-for-Artificial-Intelligence.pdf, 2018. Accessed: 2025-09-22.

NITI Aayog. Approach Document for India Part 1 – Principles for Responsible AI. https://www.niti.gov.in/sites/default/files/2021-02/Responsible-AI-22022021.pdf, 2021. Accessed: 2025-09-22.

OECD. Recommendation of the Council on Artificial Intelligence (OECD/LEGAL/0449). https://legalinstruments.oecd.org/en/instruments/OECD-LEGAL-0449, 2019. Accessed: 202509-22.

European Parliament. Artificial Intelligence Act (Regulation (EU) 2024/1689). https://eur-lex.europa.eu/legal-content/EN/TXT/?uri=CELEX%3A32024R1689, 2024.

Nitin Rane, Saurabh P Choudhary, and Jayesh Rane. Acceptance of artificial intelligence: key factors, challenges, and implementation strategies. *Journal of Applied Artificial Intelligence*, 5 (2):50–70, 2024.

Huw Roberts, Emmie Hine, Mariarosaria Taddeo, and Luciano Floridi. Global AI governance: barriers and pathways forward. *International Affairs*, 100(3):1275–1286, 2024.

Malak Sadek, Emma Kallina, Thomas Bohné, Céline Mougenot, Rafael A Calvo, and Stephen Cave. Challenges of responsible AI in practice: scoping review and recommended actions. *AI & society*, 40(1):199–215, 2025.

Lewin Schmitt. Mapping global AI governance: a nascent regime in a fragmented landscape. *AI and Ethics*, 2(2):303–314, 2022.

Nathalie A Smuha. From a 'race to AI'to a 'race to AI regulation': regulatory competition for artificial intelligence. *Law, Innovation and Technology*, 13(1):57–84, 2021.

Elham Tabassi. Artificial intelligence risk management framework (AI RMF 1.0). *NIST*, 2023.

Araz Taeihagh. Governance of artificial intelligence. *Policy and Society*, 40(2):137–157, 2021.

Telecommunication Engineering Centre. Fairness Assessment and Rating of Artificial Intelligence Systems. Technical Report TEC Standard 57050:2023, Telecommunication Engineering Centre, 2023. 55 pages.

Telecommunication Engineering Centre. Standard for the Schema and Taxonomy of an AI Incident Database in Telecommunications and Critical Digital Infrastructure. Technical Report TEC Standard 57090:2025, Telecommunication Engineering Centre, 2025. 15 pages.

The Economic Times. Government not considering regulating AI growth, says IT Minister Vaishnaw. https://economictimes.indiatimes.com/news/india/government-not-considering-regulating-ai-growth-says-it-minister-vaishnaw/articleshow/99273629.cms?from= mdr, April 2023. Accessed: 2025-09-22.

Violet Turri and Rachel Dzombak. Why we need to know more: Exploring the state of AI incident documentation practices. In *Proceedings of the 2023 AAAI/ACM Conference on AI, Ethics, and Society*, pages 576–583, 2023.

UNESCO. Recommendation on the Ethics of Artificial Intelligence. https://unesdoc.unesco.org/ark:/48223/pf0000380455, 2022. Accessed: 2025-09-22.

Katerina Yordanova and Natalie Bertels. Regulating AI: Challenges and the way forward through regulatory sandboxes. *Multidisciplinary perspectives on artificial intelligence and the law*, 441, 2024.

Liming Zhu, Xiwei Xu, Qinghua Lu, Guido Governatori, and Jon Whittle. AI and ethics—Operationalizing responsible AI. In *Humanity driven AI: Productivity, well-being, sustainability and partnership*, pages 15–33. Springer, 2021.